\documentclass[twocolumn,aps,prl,showpacs]{revtex4}

\usepackage{amsmath,epsfig,amsfonts,epsfig,graphicx,rotating
}

\begin{document}

\title{Dynamics and Manipulation of Matter-Wave Solitons in Optical Superlattices}

\author{Mason A.\ Porter$^1$, P. G.\ Kevrekidis$^2$, 
R.\ Carretero-Gonz\'{a}lez$^3$ and D. J.\ Frantzeskakis$^{4}$}
\affiliation{$^1$ Department of Physics and Center for the Physics of Information, 
California Institute of Technology, Pasadena, CA  91125, USA \\
$^2$Department of Mathematics and Statistics, University of Massachusetts, Amherst MA 01003-4515, USA \\
$^3$ Nonlinear Dynamical Systems Group (http://nlds.sdsu.edu/),%
Department of Mathematics and Statistics, San Diego State University, San Diego CA, 92182-7720, USA \\
$^4$ Department of Physics, University of Athens, Panepistimiopolis, Zografos, Athens 15784, Greece}


\begin{abstract}
We analyze the existence and stability of bright, dark, and gap matter-wave solitons 
in optical superlattices. Then, using these properties, 
we show that (time-dependent) ``dynamical superlattices'' 
can be used to controllably place, guide, and manipulate these solitons. 
In particular, we use numerical experiments to 
displace solitons by turning on a secondary lattice structure, transfer solitons from one location to another by shifting one superlattice substructure relative to the other, and implement solitonic 
``path-following'', in which a matter wave follows the time-dependent lattice 
substructure into oscillatory motion.
\end{abstract}


\maketitle


\section{Introduction}
After the first experimental realization of Bose-Einstein condensates (BECs) in dilute alkali metal vapors \cite {becna}, their study has experienced enormous experimental and theoretical advancements \cite{pethick}.  Their potential applications---ranging from matter-wave optics to precision measurements and quantum information processing---are widely held to be very promising.

External electromagnetic fields or laser beams are used to produce, trap, and manipulate BECs. 
Additionally, using highly anisotropic traps, it is possible to produce quasi one-dimensional (1D) BECs 
(see, e.g., \cite{oned}). In the formation of the latter (lying, e.g., along the $x$-direction), 
atoms are trapped using a confining magnetic or optical potential $V(x)$. 
In early experiments, only harmonic potentials were employed, but a wide variety of potentials 
can now be implemented experimentally. 
Among the most frequently studied, both experimentally 
and theoretically (see, e.g., \cite{konotopmplb} and references therein), are periodic optical 
lattice potentials created using counter-propagating laser beams \cite{olstandard}. 
Such potentials have been used to study Josephson effects \cite{anderson}, squeezed states \cite{squeeze}, Landau-Zener tunneling and Bloch oscillations \cite{morsch}, the classical \cite{smerzi} 
and quantum \cite{mott} superfluid--Mott insulator transitions, and so on.
Additionally, with each lattice site occupied by one alkali atom in its ground state, BECs in optical lattices show promise as registers for quantum computers \cite{porto}. Optical lattice potentials are, therefore, 
of particular interest from the perspective of both fundamental quantum physics and its connection to applications.

An important generalization of this setting was recently realized experimentally 
when  $^{87}$Rb atoms were loaded into an optical ``superlattice'' by the sequential 
creation of two lattice structures \cite{quasibec}. Stationary superlattices can be described mathematically in the  form
\begin{equation}
       V(x) = V_1\cos(\kappa_1 x) + V_2\cos(\kappa_2 x)  \,, \label{sl}
\end{equation}
where $\kappa_1$ and $\kappa_2 > \kappa_1$ are, respectively, the primary and 
secondary lattice wavenumbers, and $V_1$ and $V_2$ are the associated 
sublattice amplitudes.
The great flexibility of superlattice potentials arises from the fact that 
the above parameters can be tuned experimentally, providing precise control over the shape and time-variation of the external potential.  Nevertheless, despite the aforementioned experiments, there have thus far been very few theoretical studies of BECs in superlattice potentials; these include work on dark \cite{pearl1} and gap \cite{pearlsuper} solitons, the Mott-Peierls transition \cite{mottsuper}, 
non-mean-field effects \cite{anamaria}, and spatially extended solutions \cite{super,mather}.  

The aim of this work is to show that optical superlattice potentials may be used not only to sustain 
solitary matter-waves but also to manipulate them at will.  As we illustrate below, the addition 
of the secondary lattice makes optical superlattices considerably more flexible than regular optical lattices.  Using effectively 1D settings, our model is the 1D Gross-Pitaevskii (GP) equation \cite{1d} 
in the following dimensionless form \cite{konotopmplb},
\begin{equation}
	iu_t = -\frac{1}{2}u_{xx} + g|u|^2u + V(x)u \,, \label{nls}
\end{equation}
where $u$ is the mean-field BEC wavefunction, the nonlinearity coefficient $g=\pm 1$ accounts for 
repulsive and attractive interatomic interactions, respectively, and the 
potential $V(x)$ is given by Eq.~(\ref{sl}).   In this paper, we study the kinematics, stability, and dynamics of bright solitons (for $g=-1$) as well as dark and gap solitons (for $g=+1$).  We subsequently utilize ``dynamical superlattices'' (in which specific lattice parameters are time-dependent) to show that superlattice potentials can be used to controllably guide, deposit, and manipulate solitons. 
Because of this flexibility, matter-wave solitons (employed as information carriers) 
loaded into superlattice potentials may prove useful for quantum 
computing applications.


\section{Bright solitons} 
For attractive interactions ($g = -1$), and in the absence of any potential [$V(x) = 0$], 
Eq.~(\ref{nls}) possesses an exact stationary bright soliton solution of the form,
\begin{equation}
        u(x,t)= \eta\, {\rm sech}\left[\eta (x-x_0)\right]\, \exp(-i \Lambda t),
\label{solit}
\end{equation}
where $\eta$ is the amplitude (and inverse width), $\Lambda \equiv -\eta^2/2 = 
\mu$ is 
the frequency (i.e., the chemical potential), and $x_0$ is the location of the soliton center. 
Traveling solitons with a constant velocity can also be generated by applying a Galilean boost to 
the solution in Eq.~(\ref{solit}). Bright solitons have been realized in experiments \cite{expb1}, and it is 
also feasible to generate them in optical lattices and superlattices.

\begin{figure}[tbp]
\centerline{
{\epsfig{file=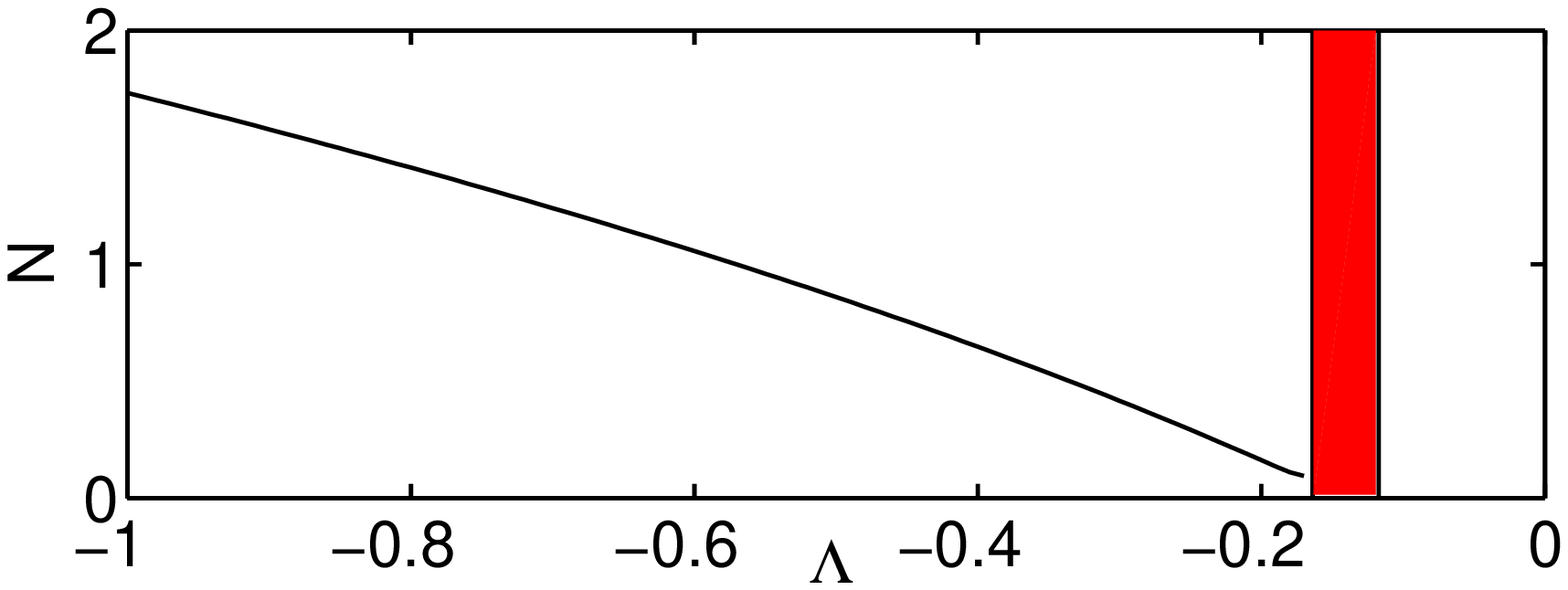, width=.417\textwidth}}}
\medskip
\centerline{
{~~\epsfig{file=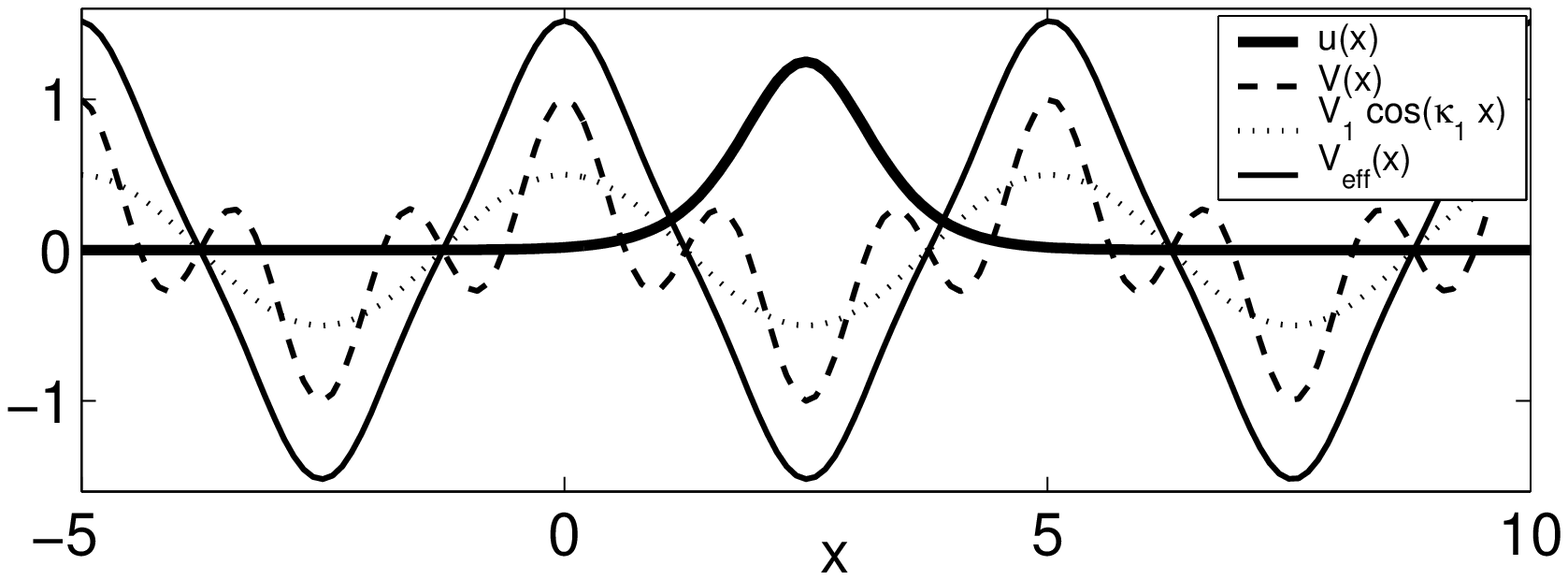, width=.4\textwidth}}}
\caption{The top panel shows the branch of stable solitons 
(centered at $x_0 \approx 2.5$), with 
the normalized number of particles $N=\int_{-\infty}^{+\infty}|u|^{2}dx$ as a function of the 
frequency $\Lambda$. The branch terminates exactly at the 
edge of the first band of excitations (the shaded rectangle) of the underlying linear spectrum.
The bottom panel shows (for $\Lambda = -1$) the profile of the solution (thick solid curve), together with the superlattice potential (dashed curve), the primary lattice potential (dotted curve), and the effective potential felt by the soliton (solid curve).} 
\label{slfig3}
\end{figure}

In the presence of the potential, Eq.~(\ref{nls}) is a perturbed 
Hamiltonian system with perturbation energy
\begin{equation}
	E_p[u]= \int_{-\infty}^{\infty} V(x) |u|^2 dx\,.  \label{P_energy}
\end{equation}
The reduced Hamiltonian $H$, obtained by inserting the solutions (\ref{solit}) 
into the perturbation energy (\ref{P_energy})
is readily evaluated to be
\cite{kivrev,scharf,kapit}
\begin{equation}
        H(x_0)= \sum_{j=1,2} \frac{\pi \kappa_j V_j }{\sinh\left(\pi \kappa_j / 2 \eta \right)} \cos(\kappa_j x_0)\,. 
\label{energy}
\end{equation}
According to Refs.~\cite{kivrev,scharf,kapit}, stationary states of the
perturbed system can be obtained by demanding that
$dH(x_0)/d x_0=0$. That is, the selected center positions $x_0$ 
are critical points of the reduced Hamiltonian. 
Following this perturbative treatment, one can also derive the equation 
of motion describing the center of the soliton as if it 
were a particle in an effective potential \cite{scharf}: 
\begin{equation}
	\frac{d^{2}x_{0}}{dt^{2}}=-\frac{1}{\eta} \frac{d H(x_0)}{d x_0} \equiv -\frac{1}{2\eta} 
\frac{d V_{\rm eff}(x_0)}{d x_0}.
\label{eff_eq}
\end{equation}
%
Furthermore, as shown rigorously in \cite{kapit}, the small, nonzero linear stability eigenvalues 
(of the linearization around the solitary wave) in the presence of the perturbation are given 
by $\lambda^2= -\eta d^2 H(x_0)/d x_0^2$. Hence, the soliton is stable (unstable) at the minima (maxima) 
of the {\it effective} potential. If initialized at a maximum, the soliton can be seen to split 
(a) symmetrically under time-evolution of Eq.~(\ref{nls}) if the two minima next to the maximum 
have the same height and (b) asymmetrically if they have different heights.



The top panel of Fig.~\ref{slfig3} shows the parametric continuation of the (always) stable 
bright soliton solution centered at the minimum of the effective potential (see the bottom panel). 
We have also obtained solutions centered at the maxima of the effective potential, but they are 
always dynamically unstable (and, hence, are not shown here). We have tried to identify solutions centered at intermediate points, as perhaps suggested by the metastable maxima and minima of the 
regular potential $V(x)$ (see the bottom panel of Fig.~\ref{slfig3}). The fact that this has not 
been possible---the effective potential lacks the metastable critical points of $V(x)$ (again, see the bottom panel of Fig.~\ref{slfig3})---confirms that the {\it effective potential}, rather than the actual one, 
is governing the steady states. Unless otherwise stated, we used the parameter 
values $V_1=V_2=0.5$ and $\kappa_2 = 3 \kappa_1=3.75$ in our numerical simulations. 
The ratio $\kappa_2/\kappa_1 = 3$ was chosen to correspond to the superlattice reported 
experimentally in Ref.~\cite{quasibec}.  We consider $V_1 = V_2$ for simplicity, 
but similar results can be obtained with other sublattice amplitudes.

The linearization of Eq.~(\ref{nls}) describes the case of a noninteracting condensate ($g = 0$). 
Its solution is given by a superposition of Bloch waves and its spectrum consists of bands of 
eigenvalues (frequencies) $\Lambda = \Lambda_{n}(k)$, where the index $n$ denotes the band index and the quasimomentum $k$ is a real wavenumber of bounded Bloch waves \cite{ashcroft}. Different bands are separated by ``gaps'' in which ${\rm Im}(k) \neq 0$. As discussed in, e.g., Refs.~{\cite{pearl1,pearlsuper}, a superlattice potential yields a more complicated linear band structure than a regular lattice, as it includes smaller gaps (called ``mini-gaps'' in \cite{pearl1,pearlsuper}) in addition to regular gaps.  With the above parameter choices, the underlying linear spectrum for the superlattice potential has its first band in the interval 
$[-0.1643,-0.1178]$ (see the shaded area in the top panel 
of Fig.~\ref{slfig3}), its second band in $[0.39,1.5244]$ (with a 
minigap in $[0.7748,0.8171]$), and so on.


\section{Dark and Gap Solitons} 
We now turn to the study of repulsive BECs with $g = 1$ and 
first consider dark soliton solutions of Eq.~(\ref{nls}). Dark solitons have 
been realized experimentally \cite{expd1} in BECs confined in harmonic traps and have 
been studied theoretically in lattices \cite{br,dsol} and superlattices \cite{pearl1}, where they can also be created.

\begin{figure}[tbp]
{\epsfig{file=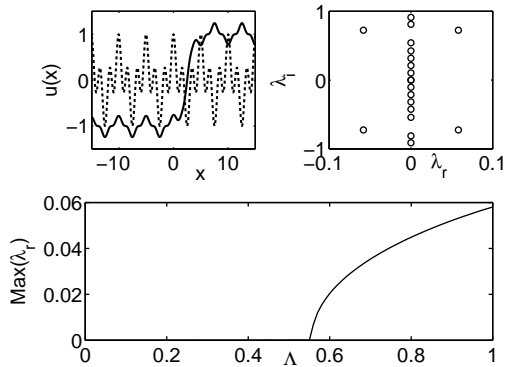, width=.4\textwidth}}
\caption{Unstable dark soliton centered at a potential minimum ($x_0 \approx 2.5$) with $\Lambda = 1$. 
The top left panel shows the stationary spatial profile $u(x)$ (solid) and the superlattice potential (dashed).  The spectral plane (top right) of the linearization eigenvalues $\lambda=\lambda_r+i\lambda_i$ reveals a so-called ``loxodromic quartet'' [i.e., all four eigenvalues have the same 
$(|{\rm Re}(\lambda)|, |{\rm Im}(\lambda)|)$] that causes the instability.  The bottom panel, showing the magnitude of the real part of the eigenvalue quartet vs. the frequency $\Lambda$, indicates that this soliton branch becomes stable for $\Lambda \lesssim 0.55$.} \label{slfig4}
\end{figure}

We seek solutions of Eq.~(\ref{nls}) of the form $u=\phi(x,t)\exp(-it)$ (we set the chemical potential $\mu=1$) and assume that the potential, characterized by a scale $R$, is slowly-varying on the soliton scale (which is on the order of the healing length).  Then, following the multiple-scale boundary layer theory of Ref.~\cite{anglin} (similar results can be obtained using other perturbative approaches \cite{imp}), we use the ansatz
\begin{equation}
	\phi = [\phi_0 + \varepsilon \phi_1(x - x_0,t)] \exp[i\bar{\theta} + i\bar{v}(x - x_0)]+ O(\varepsilon^2)
\end{equation}
for the interval $|x - x_0| < R$, where 
\begin{equation}
	\phi_0 = i(v_0 - \bar{v}) + k \tanh [k(x - x_0)]\,,
\end{equation}
with $v_0 = \dot{x}_0$ and $\bar{\theta}(t) = (1/2)[\theta(x_0 - R,t) + \theta(x_0 + R,t)]$ ($\bar{v}$ is defined analogously).  Letting the cut-off $R \rightarrow \infty$ and considering solutions satisfying $v = \partial \theta /\partial x = 0$ yields the equation of motion for the soliton center $x_{0}$,
\begin{equation}
	\frac{d^{2}x_{0}}{dt^{2}}=-\frac{1}{2} \frac{\partial V(x_0)}{\partial x_0} \equiv -\frac{\partial V_{\rm eff}}{\partial x_0}\,.
\label{eff}
\end{equation}
%

For the superlattice potential (\ref{sl}), Eq.~(\ref{eff}) becomes
\begin{equation}
\frac{d^{2}x_{0}}{dt^{2}}= \frac{1}{2}\kappa_1V_1\sin(\kappa_1x_0) 
           + \frac{1}{2}\kappa_2V_2\sin(\kappa_2x_0)\,. \label{dyn}
\end{equation}

Figure \ref{slfig4} shows a dark soliton centered at a minimum of the superlattice. 
The depicted solution is unstable due to a loxodromic quartet of eigenvalues but can become 
stable for a frequency $\Lambda \lesssim 0.55$ (see the bottom panel in Fig.~\ref{slfig4}). 
A solution centered at a maximum of the potential is unstable due to the presence of a large 
real eigenvalue pair (e.g., for $\Lambda=1$, the eigenvalues have magnitude $0.3833$). 
We have also attempted to identify solutions centered between consecutive absolute maxima and minima, i.e., near metastable minima and maxima of the effective potential (e.g., for $x \approx 0.9$ 
and $x \approx1.6$), but were unable to find any. This apparent
violation of the 
effective energy landscape suggested by the boundary layer theory
poses an interesting question for future studies.



In the same context of repulsive condensates (again with $g = 1$), we have also obtained gap soliton solutions of Eq.~(\ref{nls}).  Such gap solitons are spatially localized nonlinear modes that occur in the band gaps of the linearized spectrum and, in fact, have the form of {\em bright} solitons but in a repulsive medium.  Figure \ref{slfig5} shows a stable branch of gap solitons that exists throughout the entire finite gap between the first and second bands. Gap solitons have recently been obtained experimentally 
in regular optical lattices \cite{markus}, and thus they can also be straightforwardly produced experimentally in the superlattice setting.

\begin{figure}[tbp]
{\epsfig{file=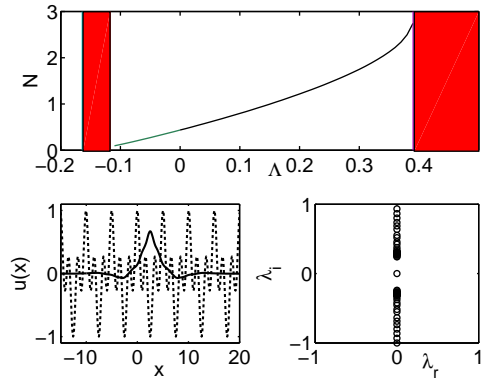, width=.4\textwidth}}
\caption{A branch of gap solitons in the finite gap between the first and second bands.  The top panel
shows the normalized number of particles $N$ in each soliton and the lower two panels show a 
typical example of the solution (around the middle of the branch) for $\Lambda=0.15$.  
The stability of the solitons depends on the value of the chemical potential
(and may also be affected for finite domains by the size of the computational
domain).} 
\label{slfig5}
\end{figure}


\begin{figure}[tbp]
{\epsfig{file=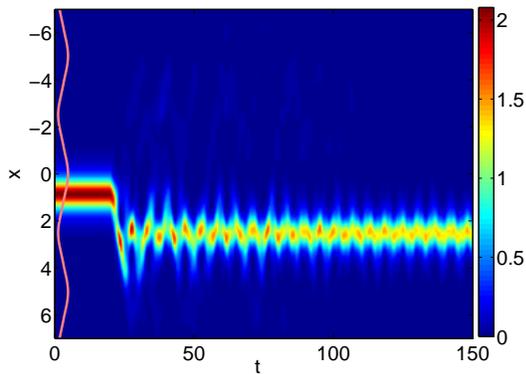, width=.4\textwidth}}

\caption{(Color Online).
Bright soliton evolution for the $V_2\cos(\kappa_2 x)$ 
potential. The soliton is initially centered at the minimum near $0.8$, which becomes a 
metastable minimum of the superlattice potential $V(x)$ and disappears 
altogether as a minimum for the effective potential $V_{\rm eff}(x)$ once the 
$V_1\cos(\kappa_1 x)$ potential is turned on.  The abrupt switching-on of the
second sublattice at $t = 20$ is represented functionally by
$V_1(t)=(1/2)[1+\tanh(5(t-20))]$. The soliton can no longer stay in its 
original location, so it goes to the closest minimum of the effective
potential (solid red/light curve), about which it oscillates.} 
\label{slfig6}
\end{figure}

\section{Dynamical Superlattices} 
We now turn to using the superlattice as a means to guide, displace, and (more generally) manipulate matter-wave solitons at will in the 
potential landscapes discussed above. In recent experiments \cite{chu}, the center of a regular optical 
lattice was ``shaken'' (translated periodically) to examine a period-doubling instability in BECs.  Such dynamical manipulations of superlattice potentials are similarly achievable in the laboratory.

Given a stable bright soliton located at a minimum of a regular optical lattice, we can displace it from 
its location by turning on the second sublattice abruptly (nonadiabatically), as indicated by the space-time plot in Fig.~\ref{slfig6}. As the soliton is no longer located at a minimum of the effective potential, it cannot stay in its original location, so it moves to the closest minimum and oscillates in that well.

\begin{figure}[tbp]
\centerline{Bright soliton transfer}
\medskip
\centerline{
{\epsfig{file=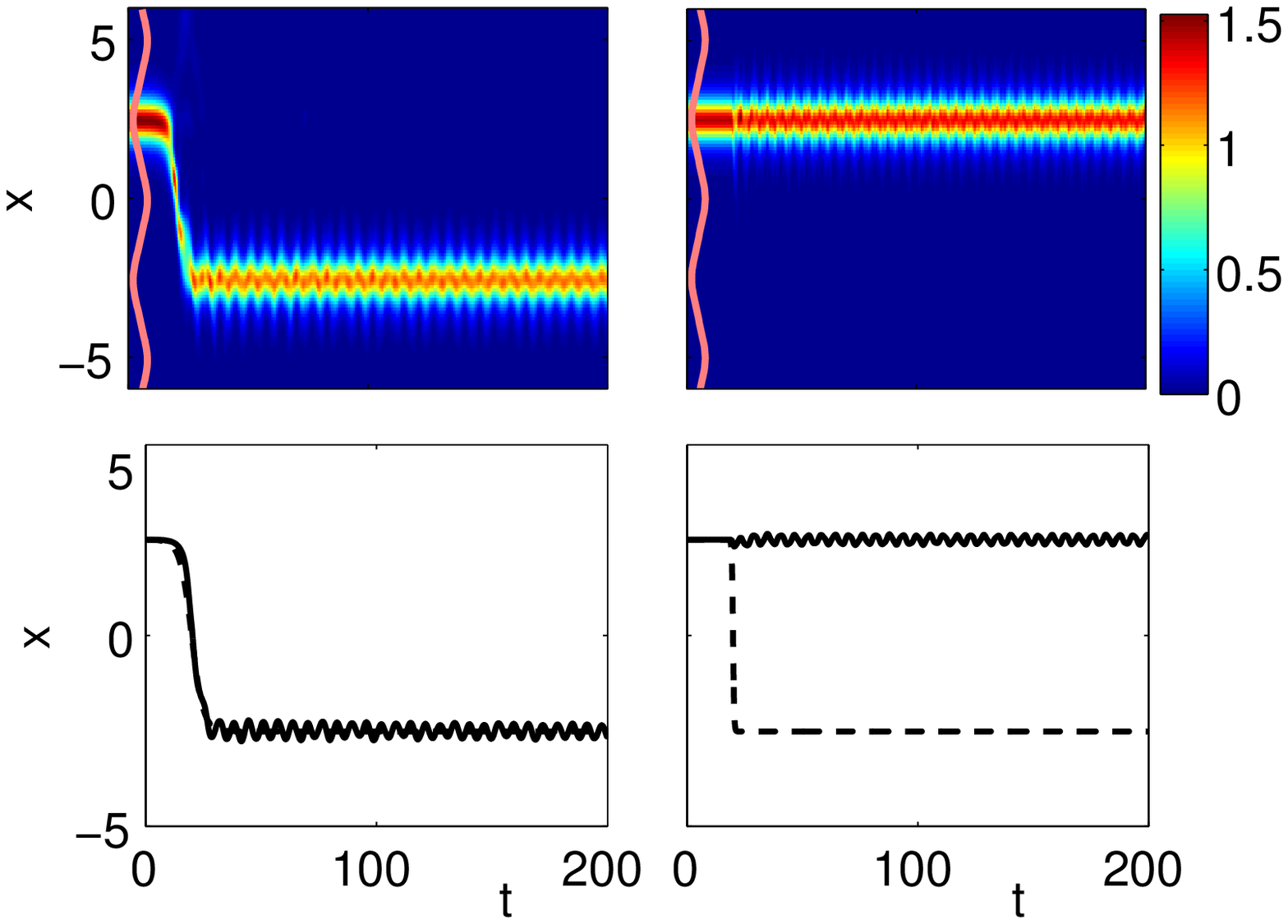,  width=.48\textwidth, height=5.2cm}}}
\medskip
\medskip
\centerline{Gap soliton transfer}
\medskip
\centerline{
{\epsfig{file=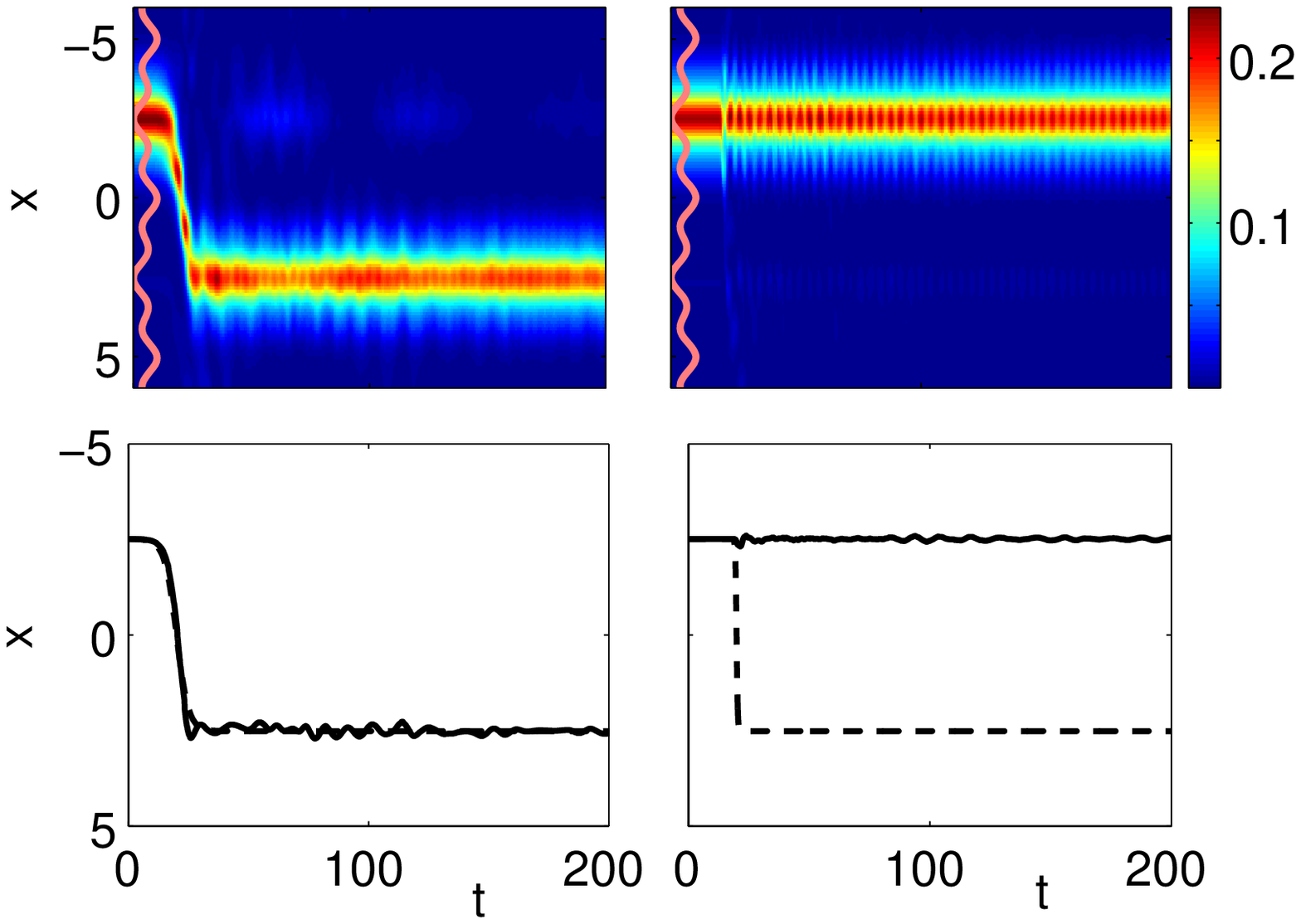,  width=.48\textwidth, height=5.2cm}}}
\caption{(Color Online). Controllable transfer of bright (top quartet) and gap 
(bottom quartet) solitons by manipulating one sublattice of the superlattice potential. 
For each quartet, the top panels show space-time plots (with colors indicating the value of $|u|^2$) 
and the bottom panels show the time-evolution of the soliton center (solid) and superlattice ``center'' $x_c$ (dashed).  The left panels show the result of adiabatic potential displacement ($\tau = 5$) and the right panels show the result of nonadiabatic potential displacement ($\tau = 0.5$). 
Clearly, adiabaticity plays a major role in the success of the soliton transfer.  
}\label{slfig7}
\end{figure}

As another example, in Fig.~\ref{slfig7}, we show how to use a sublattice of the superlattice as a means of transferring bright and gap solitons at will. A soliton is placed in the potential $V_1\cos(\kappa_1(x-x_c)) + V_2 \cos(\kappa_2 (x))$, the first sublattice of which is subsequently displaced according to the following equation,
%
\begin{equation}
	x_c=x_{{i}}+\frac{1}{2}(x_{{f}}-x_{{i}})\left[1+\tanh \left( \frac{t-t_0}{\tau} \right) \right]\,,
\end{equation}	
where $x_{i}$ and $x_{f}$  denote, respectively, the initial and final ``center'' positions 
and the parameter $\tau$ determines the speed of the displacement \cite{dsol,hector}.  
We considered both adiabatic cases (large $\tau$; left panels), where the soliton transfer can be successful, and nonadiabatic ones (small $\tau$; right panels), 
where such manipulations will typically fail to guide the soliton.  Note that similar results have also been obtained both for bright \cite{prabr} and dark \cite{dsol} solitons but in regular optical lattices.

We also performed a more ``demanding'' experiment, in which the bright and gap solitons were not merely deposited at a new location but were instead ``instructed'' to follow the time-dependent sublattice into oscillatory motion.  This numerical experiment, shown in Fig.~\ref{slfig9}, corresponds to a lattice
displaced according to 
\begin{equation}
	x_c=x_{{i}}+ \frac{x_{f}}{2} \sin \left(\frac{t-t_0}{\tau_1}\right)
\left[1+\tanh \left(\frac{t-t_0}{\tau}\right)\right]\,.
\label{new}
\end{equation}	
We turn on the sinusoidal potential abruptly (with $\tau=0.5$), but vary 
its period (using both large and small $\tau_1$). We observe that while 
the solitons lose power (i.e., $\int dx |u|^2$)
in traversing the rough terrain of the immobile lattice (in this example, the bright soliton with the larger power emits more radiation), they can still follow the oscillation, provided the 
motion is, again, sufficiently adiabatic. For small $\tau_1$ (the nonadiabatic case), the waves completely disintegrate into small-amplitude radiation.


\begin{figure}[tbp]
\centerline{Bright soliton path-following}
\medskip
\centerline{
{\epsfig{file=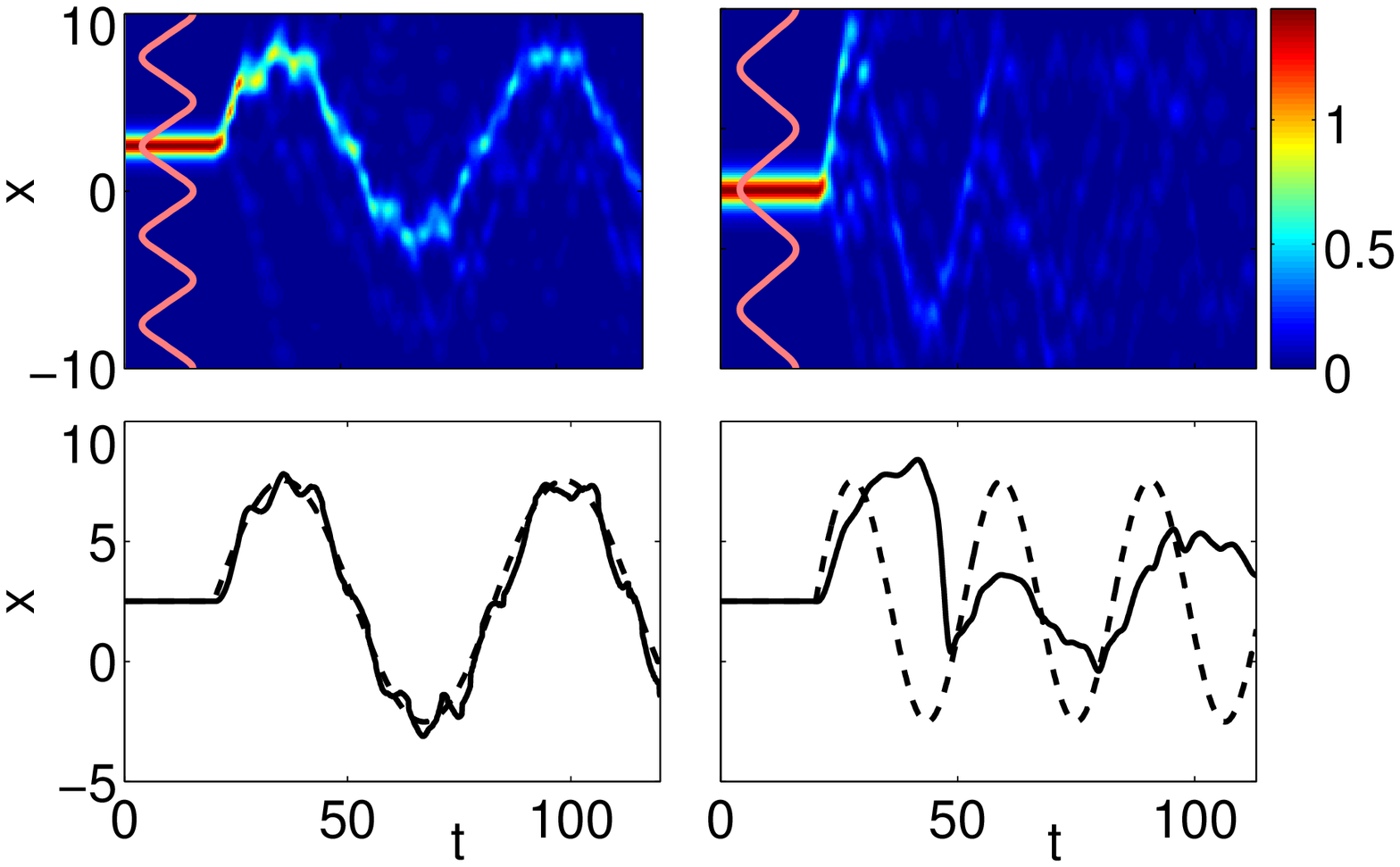,  width=.48\textwidth, height=5.2cm}}} 
\medskip
\medskip
\centerline{Gap soliton path-following}
\medskip
\centerline{
{\epsfig{file=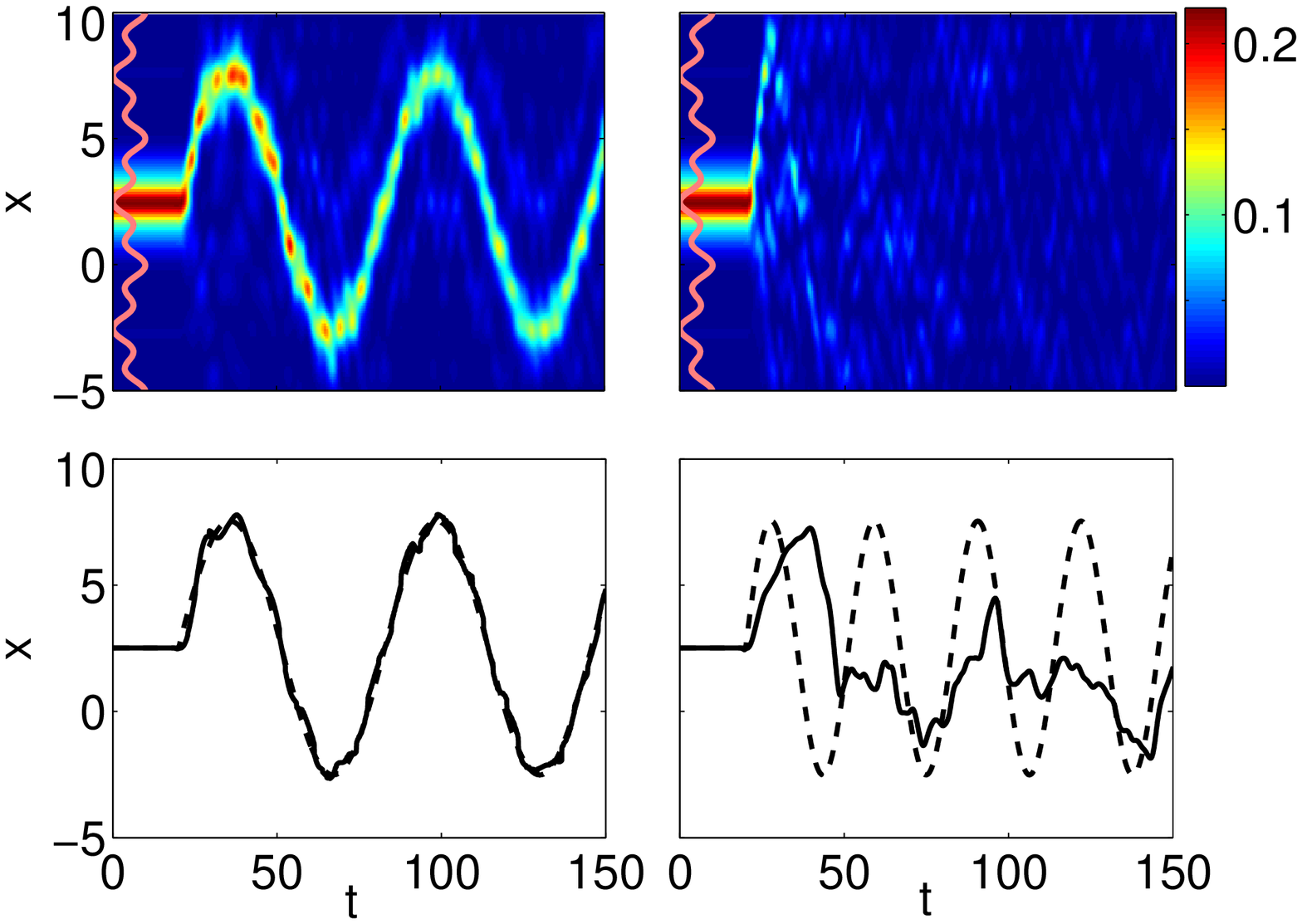, width=.48\textwidth, height=5.2cm}}}
\caption{(Color Online).  As in Fig.~\ref{slfig7}, but for controllable path-following of bright 
(top quartet) and gap (bottom quartet) solitons. The evolution of the soliton center is governed 
by Eq.~(\ref{new}) for $\tau_1=10$ (left panels) and $\tau_1=5$ (right panels) [$\tau=0.5$ in both cases]. 
Clearly, it is necessary that the lattice oscillations be sufficiently adiabatic for solitons to follow 
a dynamical superlattice's path successfully.
} \label{slfig9}
\end{figure}




\section{Conclusions} 
In this work, we analyzed the properties of bright, dark, and gap matter-wave solitons 
in the presence of superlattice potentials. We focused, in particular, 
on showing (for typical potential parameter values) how the dynamical modification of the (experimentally tunable) properties of the 
superlattice potential can be used to guide, transfer, and deposit these
coherent structures across their corresponding energy landscape.
We believe that such controllable soliton manipulation benefits greatly from the enhanced flexibility 
arising from the extra length scale and tunable parameters of the superlattice structure as 
compared to regular optical lattices. This, in turn, 
may pave the way towards the macroscopic
manipulation of solitonic ``bits'' of information in 
BECs in a manner that bears similarities to 
(but also enables extensions of) the setting of nonlinear optics.



We thank T. Kapitula and R. Hulet for numerous useful discussions and gratefully acknowledge support 
from the Gordon and Betty Moore Foundation (M.A.P.), NSF-DMS-0204585, NSF-CAREER (P.G.K.), and NSF-DMS-0505663 (P.G.K. and R.C.G.).



\begin{thebibliography}{99}

\bibitem{becna} 
M. H. Anderson, J. R. Ensher, M. R. Matthews, C. E. Wieman, and E. A. Cornell, Science {\bf 269}, 198 (1995);
K. B. Davis, M.-O. Mewes, M. R. Andrews, N. J. van Druten, 
D. S. Durfee, D. M. Kurn, and W. Ketterle, Phys. Rev. Lett. {\bf 75}, 3969 (1995);
C. C. Bradley, C. A. Sackett, J. J. Tollett, and R. G. Hulet, Phys. Rev. Lett. {\bf 75}, 1687 (1995).

\bibitem{pethick} 
F.\ Dalfovo, S. Giorgini, L. Pitaevskii, and S. Stringari, Rev.\ Mod.\ Phys.\ {\bf 71}, 463 (1999); 
C. J.\ Pethick and H.\ Smith, {\it Bose-Einstein Condensation in Dilute Gases} Cambridge University Press (2002).

\bibitem{oned} 
A. G{\"o}rlitz, J. M. Vogels, A. E. Leanhardt, C. Raman, T. L. Gustavson, J. R. Abo-Shaeer, 
A. P. Chikkatur, S. Gupta, S. Inouye, T. Rosenband, and W. Ketterle, Phys. Rev. Lett. {\bf 87}, 130402 (2001); 
F. Schreck, L. Khaykovich, K. L. Corwin, G. Ferrari, T. Bourdel, J. Cubizolles, and C. Salomon, Phys. Rev. Lett. {\bf 87}, 080403 (2001); 
M. Greiner, I. Bloch, O. Mandel, T. W. H\"ansch, and T. Esslinger, Phys. Rev. Lett. {\bf 87}, 160405 (2001);

\bibitem{konotopmplb} 
P. G.\ Kevrekidis and D. J.\ Frantzeskakis, Mod.\ Phys.\ Lett.\ B \textbf{18}, 173 (2004);
V. A.\ Brazhnyi and V. V.\ Konotop, Mod.\ Phys.\ Lett.\ B \textbf{18}, 627 (2004).

\bibitem{olstandard} 
J. H. Denschlag, J. E. Simsarian, H. H\"affner, C. McKenzie, A. Browaeys, D. Cho, K. Helmerson, 
S. L. Rolston, and W. D. Phillips, J. Phys. B {\bf 35}, 3095 (2002).

\bibitem{anderson} 
B. P.\ Anderson and M.A.\ Kasevich, Science {\bf 282}, 1686 (1998).

\bibitem{squeeze} 
C. Orzel, A. K. Tuchman, M. L. Fenselau, M. Yasuda, and M. A. Kasevich, Science {\bf 291}, 2386 (2001).

\bibitem{morsch} 
O. Morsch, J. H. M\"uller, M. Cristiani, D. Ciampini, and E. Arimondo, Phys. Rev. Lett. {\bf 87}, 140402 (2001);
M. Jona-Lasinio, O. Morsch, M. Cristiani, N. Malossi, 
J.H. M\"uller, E. Courtade, M. Anderlini, and E. Arimondo, 
Phys. Rev. Lett. {\bf 91}, 230406 (2003);
V.V. Konotop, P.G. Kevrekidis and M. Salerno,
Phys. Rev. A {\bf 72}, 023611 (2005).

\bibitem{smerzi} 
A. Smerzi, A. Trombettoni, P. G. Kevrekidis, and A. R. Bishop, Phys. Rev. Lett. {\bf 89}, 170402 (2002); 
F. S. Cataliotti, L. Fallani, F. Ferlaino, C. Fort, P. Maddaloni, M. Inguscio, New J. Phys. {\bf 5}, 71 (2003).

\bibitem{mott} 
M. Greiner, O. Mandel, T. Esslinger, T. W. H\"ansch, I. Bloch, Nature {\bf 415}, 39 (2002).

\bibitem{porto} 
J. V. Porto, S. Rolston, B. L. Tolra, C. J. Williams, W. D. Phillips, Phil. Trans.: Math. Phys. Eng. Sci. {\bf 361}, 1471 (2003); 
K. G. H. Vollbrecht, E. Solano, and J. I. Cirac, Phys. Rev. Lett. {\bf 93}, 220502 (2004).





\bibitem{quasibec} 
S. Peil, J. V. Porto, B. L. Tolra, J. M. Obrecht, B. E. King, 
M. Subbotin, S. L. Rolston, and W. D. Phillips, Phys. Rev. A {\bf 67}, 051603(R) (2003).

\bibitem{pearl1} 
P. J. Y. Louis, E. A. Ostrovskaya, and Y. S. Kivshar, J. Opt. B {\bf 6}, S309 (2004).

\bibitem{pearlsuper} 
P. J. Y. Louis, E. A. Ostrovskaya, and Y. S. Kivshar, Phys. Rev. A {\bf 71}, 023612 (2005).

\bibitem{mottsuper} 
L. A. Dmitrieva and Y. A. Kuperin, cond-mat/0311468.

\bibitem{anamaria} 
A. M. Rey, B. L. Hu, E. Calzetta, A. Roura, and C. W. Clark, Phys. Rev. A {\bf 69}, 033610 (2004).

\bibitem{super} M. A. Porter and P. G. Kevrekidis, SIAM J. App. Dyn. Sys. {\bf 4}, 4, 783 (2005).

\bibitem{mather} 
M. van Noort, M. A. Porter, Y. Yi, S.-N. Chow, math.DS/0405112;
V. P. Chua and M. A. Porter, Int. J. Bif. Chaos (in press).


\bibitem{1d} 
V. M.\ P\'{e}rez-Garc\'{\i}a, H. Michinel, H. Herrero, Phys.\ Rev.\ A \textbf{57}, 3837 (1998).


\bibitem{expb1}  
K. E. Strecker, G. B. Partridge, A. G. Truscott, and R. G. Hulet, Nature {\bf 417}, 150 (2002);
L. Khaykovich, F. Schreck, G. Ferrari, T. Bourdel, J. Cubizolles, L. Carr, Y. Castin, 
and C. Salomon, Science {\bf 296}, 1290 (2002).



\bibitem{kivrev} 
Y. S. Kivshar and B. A. Malomed, Rev. Mod. Phys. {\bf 61}, 763 (1989).

\bibitem{scharf} 
R.\ Scharf and A. R.\ Bishop, Phys.\ Rev.\ E {\bf 47}, 1375 (1993).

\bibitem{kapit} T. Kapitula, Physica D {\bf 156}, 186 (2001).

\bibitem{ashcroft} N. W. Ashcroft and N. D. Mermin, {\it Solid State Physics}, Brooks/Cole (1976).


\bibitem{expd1} 
S. Burger, K. Bongs, S. Dettmer, W. Ertmer, K. Sengstock, 
A. Sanpera, G. V. Shlyapnikov, and M. Lewenstein, Phys. Rev. Lett. {\bf 83}, 5198 (1999); 
J. H. Denschlag, , J. E. Simsarian, D. L. Feder, C. W. Clark, L. A. Collins, J. Cubizolles, 
L. Deng, E. W. Hagley, K. Helmerson, W. P. Reinhardt, S. L. Rolston, B. I. Schneider, 
and W. D. Phillips, Science {\bf 287}, 97 (2000).

\bibitem{br} 
P. G. Kevrekidis, R. Carretero-Gonz\'alez, G. Theocharis, 
D. J. Frantzeskakis, and B. A. Malomed, Phys. Rev. A {\bf 68}, 035602 (2003).

\bibitem{dsol} G. Theocharis, D. J. Frantzeskakis, R. Carretero-Gonz\'alez, P. G. Kevrekidis, and B. A. Malomed, 
Phys. Rev. E {\bf 71}, 017602 (2005).

\bibitem{anglin} T.\ Busch and J. R.\ Anglin, Phys.\ Rev.\ Lett.\ {\bf 84}, 2298 (2000).

\bibitem{imp} 
D. J.\ Frantzeskakis, G. Theocharis, F. K. Diakonos, P. Schmelcher, 
and Y. S. Kivshar, Phys.\ Rev.\ A {\bf 66}, 053608 (2002); 
D. E. Pelinovsky, D. J. Frantzeskakis, and P. G. Kevrekidis, Phys. Rev. E {\bf 72}, 016615 (2005).


\bibitem{markus} B. Eiermann, Th. Anker, M. Albiez, M. Taglieber, P. Treutlein, 
K.-P. Marzlin, and M. K. Oberthaler, Phys. Rev. Lett. {\bf 92}, 230401 (2004).

\bibitem{chu} N. Gemelke, E. Sarajlic, Y. Bidel, S. Hong, S. Chu, cond-mat/0504311.

\bibitem{hector} H. E. Nistazakis, P. G. Kevrekidis, B. A. Malomed, D. J. Frantzeskakis, and A. R. Bishop, 
Phys. Rev. E {\bf 66}, 015601 (2002).

\bibitem{prabr} P. G. Kevrekidis, D. J. Frantzeskakis, R. Carretero-Gonz\'alez, B. A. Malomed, G. Herring, and A. R. Bishop, 
Phys. Rev. A {\bf 71}, 023614 (2005).

\end{thebibliography}

\end{document}